\documentclass[aps,prd,superscriptaddress,twocolumn,showpacs]{revtex4}
\usepackage{graphicx}
\usepackage{epstopdf}
\usepackage{amsmath}
\usepackage{amsfonts}
\usepackage{amssymb}
\usepackage{latexsym}

\begin{document}
\title{Remarks on the static potential in theories with Lorentz violation terms}
\author{Patricio Gaete} \email{patricio.gaete@usm.cl} 
\affiliation{Departamento de F\'{i}sica and Centro Cient\'{i}fico-Tecnol\'ogico de Valpara\'{i}so, Universidad
T\'{e}cnica Federico Santa Mar\'{i}a, Valpara\'{i}so, Chile}
\author{Jos\'{e} A. Helay\"{e}l-Neto}\email{helayel@cbpf.br}
\affiliation{Centro Brasileiro de Pesquisas F\'{i}sicas (CBPF), Rio de Janeiro, RJ, Brasil} 
\date{\today}

\begin{abstract}
We study the impact of Lorentz violating terms on a physical observable for both electrodynamics of chiral 
matter and an Abelian Higgs-like model in $3+1$ dimensions. Our calculation is done within the 
framework of the gauge-invariant, but path-dependent, variables formalism. Interestingly
enough, for electrodynamics of chiral matter we obtain a logarithmic correction to the usual static
Coulomb potential. Whereas for a Abelian Higgs model with a Lorentz-breaking term, our result 
displays new corrections to the Yukawa potential.
\end{abstract}
\maketitle

\section{Introduction}

The study of the physical consequences of topology or, more precisely, topological terms have 
considerably increased over the last years. In all these studies, the $\theta$-term (or axion-term) $\sim
\theta {\bf E} \cdot {\bf B}$, has been the focus of interest.  As is well known, the axion is a hypothetical
pseudo-scalar particle introduced to explain the CP nonviolation problem in QCD \cite{Peccei-Quinn,Weinberg,Wilczek1}. In this respect, we also recall that the axion term provides a consistent
framework for the Witten effect \cite{Witten,Rosenberg,Franz} as well as  for the topological magneto-electric
effect \cite{Qi,Essin}. Particularly impressive is that the effects of this topological term have been 
materialized through the discovery of new materials \cite{Kane}.

On the other hand, in recent times a great deal of attention has been devoted to the study of quantum-anomaly induced effects with chiral fermions \cite{Cao}. An example that illustrates this is the chiral magnetic effect (CME), which is the generation of vector current by an external magnetic field in the presence of imbalance between the chemical potentials of right-handed and left-handed fermions \cite{Kharzeev1,Kharzeev2,Tuchin,Hayata}. Along the same line, we also mention the anomalous Hall effect \cite{Haldane,Huang}. However, it should be emphasized that the crucial feature of these quantum-anomaly induced effects is to change the electromagnetic response of chiral matter. Interestingly, these systems (electrodynamics of chiral matter) are described by a Maxwell-Chern-Simons electrodynamics with a constant four-vector, which breaks the Lorentz invariance. Let us also mention here that the issue of Lorentz invariance violation in quantum field theories has been a subject of intense study \cite{Jackiw,Kostelecky, Liberati, VAKostelecky}, where the most studied framework is the standard model extension, which consists of the minimal standard model plus small Lorentz violating terms. Particularly significant from this point of view are the Lorentz invariance violation electrodynamics, including either even or odd violating terms. 
It is worth recalling at this stage that theories with a topological term in $(2+1)$ dimensions, where the physical excitations obeying it are called anyons, have been widely discussed in the literature \cite{Deser, Dunne,Khare,Banerjee}. Accordingly, the three-dimensional Chern-Simons gauge theory offers a natural setting so that Wilczek's charge-flux composite model of the anyon can be realized \cite{Wilczek2}. 

In this context we also point out that an Abelian Higgs model with a Lorentz-breaking term has been considered in Ref. \cite{Baeta,Casana1,Casana2}, where aspects of causality, unitarity, spontaneous gauge-symmetry breaking, and vortex formation were investigated. As a result, it was shown that unitarity is always violated for an external vector time-like or null. However, whenever the external vector is space-like, physically consistent excitations are found. Also, it was found a physical feature analogous of what happens in $(2+1)$D, namely, the electrostatic and magnetostaic fields are not independent. 

With these ideas in mind, in this work we examine another aspect of these theories, that is, the impact of the Lorentz violating terms on a physical observable. To this end we will study the confinement versus screening issue for both  electrodynamics of chiral matter and an Abelian Higgs model with a Lorentz-breaking term. Our calculation is accomplished by making use of the gauge-invariant, but path-dependent, variables formalism along the lines of \cite{Pato1,Pato2}.
As we shall see, in the case of electrodynamics of chiral matter, by adopting a purely space-like vector $v^{\mu}$, 
our result shows that the static potential is a logarithmic correction to the usual static Coulomb potential. On the other hand, in the case of the Abelian Higgs model with a Lorentz-breaking term and a purely space-like vector $v^{\mu}$, the static potential displays new corrections to the Yukawa potential.

\section{Interaction energy}

\subsection{Electrodynamics of chiral matter}

We turn now  to the problem of obtaining the interaction energy between static point-like sources for the two models we shall consider in this work. With this purpose, let us consider first the Hamiltonian analysis for the electrodynamics of chiral matter coupled to an external source $J^{0}$. We start from the four-dimensional spacetime Lagrangian density 
\begin{equation}
{\cal L} = - \frac{1}{4}{F_{\mu \nu }}{F^{\mu \nu }} + \frac{\mu }{4}\theta {\tilde F^{\mu \nu }}{F_{\mu \nu }}-{A_0 }{J^0 }, \label{qED05}
\end{equation}
where $\theta  = \theta \left( {t,{\bf x}} \right)$.

Note that the Lagrangian (\ref{qED05}) can be written alternatively in the form
\begin{equation}
{\cal L} = - \frac{1}{4}{F_{\mu \nu }}{F^{\mu \nu }} - \frac{\mu }{4}{\partial _\mu }\theta {\varepsilon ^{\mu \nu \rho \sigma }}{A_\nu }{F_{\rho \sigma }}-{A_0 }{J^0 }. \label{qED10}
\end{equation}
Letting ${v_\mu } = {\partial _\mu }\theta$, we can, therefore, write
\begin{equation}
{\cal L} = - \frac{1}{4}{F_{\mu \nu }}{F^{\mu \nu }} - \frac{\mu }{4} {v_\mu }{\varepsilon ^{\mu \nu \rho \sigma }}{A_\nu }{F_{\rho \sigma }}-{A_0 }{J^0 }. \label{qED15}
\end{equation}
It may be noted here that the choice of the background four-vector ${v}_{\mathit{\mu}}$ as the four-gradient of the scalar $\theta$ automatically ensures gauge-invariance of the Carroll-Field-Jackiw term. Moreover, in a supersymmetric scenario with Lorentz-symmetry violation, supersymmetry imposes that ${v}_{\mathit{\mu}}$ must necessarily be a gradient. And, this does not mean that the background becomes dynamical: the fact that ${v}_{\mathit{\mu}}$ stems from a scalar does not impose that a ${v}_{\mathit{\mu}}$-kinetic term must be introduced into the Carroll-Field-Jackiw Lagrangian.

Having characterized the new effective Lagrangian, we can now examine the Hamiltonian structure of the theory under consideration when ${v_\mu } = \left( {0,{v_i}} \right)$. The canonical momenta is
\begin{equation}
{\Pi ^\lambda } = {F^{\lambda 0}} + \frac{\mu }{2}{\varepsilon ^{i\nu 0\lambda }}{v_i}{A_\nu },  \label{qED20}
\end{equation}
which results in the usual primary constraint, $\Pi ^0=0$ and ${\Pi ^i} = {F^{i0}} + \frac{\mu }{2}{\varepsilon ^{0ijk}}{v_j}{A_k}$. This allows us to write the following canonical Hamiltonian $H_C$:
\begin{eqnarray}
{H_C} &=& \int {{d^3}x} \left\{ { - \frac{1}{2}{F_{i0}}{F^{i0}} + \frac{1}{4}{F_{ij}}{F^{ij}}} \right\} \nonumber\\
 &+& \int {{d^3}x} \left\{ { - {A_0}\left( {{\partial _i}{\Pi ^i} - \frac{\mu }{2}{\varepsilon ^{0ijk}}{v_i}{\partial _j}{A_k} + {J^0}} \right)} \right\}. \nonumber\\
\label{qED25}
\end{eqnarray}
The secondary constraint generated by the time preservation of the primary constraint, $\Pi ^0=0$, is now
${\Gamma _1} \equiv {\partial _i}{\Pi ^i} - \frac{\mu }{2}{\varepsilon ^{0ijk}}{v_i}{\partial _j}{A_k} + {J^0} = 0$. The above constraints are the first-class constraints of the theory since no more constraints are generated by the preservation of the secondary constraint. The corresponding total (first-class) Hamiltonian  that generates the time evolution of the dynamical variables then reads $H = H_C  + \int {d^3 x} \left( {u_0(x) \Pi_0(x)  + u_1(x) \Gamma _1(x) } \right)$, where $u_o(x)$ and $u_1(x)$ are arbitrary Lagrange multipliers to implement the constraints. Since $\Pi^0=0$ always and ${\dot {A_0}}\left( x \right) = \left[ {{A_0}\left( x \right),H} \right] = {u_0}\left( x \right)$, which is completely arbitrary, we eliminate $A^0$ and $\Pi^0$ because they add nothing to the description of the system. The Hamiltonian then takes the form
\begin{eqnarray}
H &=& \int {{d^3}x} \left\{ { - \frac{1}{2}{F_{i0}}{F^{i0}} + \frac{1}{4}{F_{ij}}{F^{ij}}} \right\} \nonumber\\
 &+& \int {{d^3}x} \left\{ { w(x)\left( {{\partial _i}{\Pi ^i} - \frac{\mu }{2}{\varepsilon ^{0ijk}}{v_i}{\partial _j}{A_k} + {J^0}} \right)} \right\}, \nonumber\\
\label{qED30}
\end{eqnarray}
where $w(x) = u_1 (x) - A_0 (x)$.

In order to break the gauge freedom of the theory, we introduce a gauge condition such that the full set of constraints becomes second class, so we choose
\begin{equation}
\Gamma _2 \left( x \right) \equiv \int\limits_{C_{\xi x} } {dz^\nu
} A_\nu \left( z \right) \equiv \int\limits_0^1 {d\lambda x^i }
A_i \left( {\lambda x} \right) = 0. \label{qED35}
\end{equation}
where  $\lambda$ $(0\leq \lambda\leq1)$ is the parameter describing
the space-like straight path $ x^i = \xi ^i  + \lambda \left( {x -
\xi } \right)^i $, and $ \xi $ is a fixed point (reference point).
There is no essential loss of generality if we restrict our
considerations to $ \xi ^i=0 $. With this, the only non-trivial Dirac bracket is given by
\begin{eqnarray}
\left\{ {A_i \left( {\bf x} \right),\Pi ^j \left( {\bf y} \right)} \right\}^ *
&=& \delta _i^j \delta ^{\left( 3 \right)} \left( {{\bf x} - {\bf y}} \right) \nonumber\\
&-& \partial _i^x \int\limits_0^1 {d\lambda x^i } \delta ^{\left( 3
\right)} \left( {\lambda {\bf x} - {\bf y}} \right). \label{qED40}
\end{eqnarray}

From this expression we readily obtain the Dirac brackets in terms of the magnetic (${B^i} = {\varepsilon ^{ijk}}{\partial _j}{A_k} $) and electric (${E^i} = {\Pi ^i} - \frac{\mu }{2}{\varepsilon ^{0ijk}}{v_j}{A_k}$) fields as:
\begin{equation}
{\left\{ {{E_i}(x),{B_j}(y)} \right\}^ * } = {\varepsilon _{ijk}}{\partial ^k}{\delta ^{\left( 3 \right)}}\left( {{\bf x} - {\bf y}} \right), \label{qED45}
\end{equation}
\begin{equation}
{\left\{ {B\left( x \right),B\left( y \right)} \right\}^ * } = 0, \label{qED50}
\end{equation}
and
\begin{equation}
{\left\{ {{E_i}(x),{E_j}(y)} \right\}^ * } = \mu {\varepsilon _{ijk}}{v^k}{\delta ^{\left( 3 \right)}}\left( {{\bf x} - {\bf y}} \right). \label{qED55}
\end{equation}
This allows us to derive the equations of motion for the electric and magnetic fields, that is,
\begin{equation}
{\dot B_i}\left( x \right) = {\varepsilon _{ijk}}{\partial _k}{E_j}\left( x \right), \label{qED60}
\end{equation}
and
\begin{equation}
{\dot E_i}\left( x \right) =  - \mu {\varepsilon _{ijk}}{v_k}{E_j}\left( x \right) + {\varepsilon _{ijk}}{\partial _k}{B_j}\left( x \right). \label{qED65}
\end{equation}
Similarly, we see that Gauss's law takes the form
\begin{equation}
\frac{{\left( {{\nabla ^2} + {\mu ^2}{{\bf v}^2}} \right)}}{{{\nabla ^2}}}{\partial _i}{E^i} + {\mu ^2}\frac{{{v_j}{\partial ^j}{v_i}{E^i}}}{{{\nabla ^2}}} =  - {J^0}.\label{qED70}
\end{equation}

We also note that under the assumed conditions of static fields, equations (\ref{qED60}) and (\ref{qED65}) must vanish, which, then, yields 
\begin{equation}
{E_i} = {\partial _i}\Phi, \label{qED75}
\end{equation}
where
\begin{equation}
\Phi  = \frac{{{\nabla ^2}}}{{\left[ {{\nabla ^4} + {\mu ^2}{{\bf v}^2}{\nabla ^2} - {\mu ^2}{{\left( {{\bf v} \cdot \nabla } \right)}^2}} \right]}}\left( { - {J^0}} \right). \label{qED75-b}
\end{equation}

{For, ${J^0}\left( {\bf x} \right) = q{\delta ^{\left( 3 \right)}}\left( {\bf x} \right)$, it follows that
\begin{equation}
\Phi{=}{q}\int{\frac{{d}^{3}k}{\left({{2}\mathit{\pi}}\right)}}\frac{{{\bf k}}^{2}}{\left[{{{\bf k}}^{4}{-}{\mathit{\mu}}^{2}{\bf v}^{2}{\bf k}^{2}{+}{\mathit{\mu}}^{2}{\left({{\bf v}\cdot{\bf k}}\right)}^{2}}\right]}{e}^{{i{\bf k}}\cdot{\bf x}}. \label{qED75-c}
\end{equation}

Now, considering ${\bf v}{=}{v}\hat{z}$, we find that the foregoing equation can be brought to the form
\begin{eqnarray}
{\Phi{=}\frac{q}{{2}\mathit{\mu}{v}}\int{\frac{{d{\bf k}}_{\bot}}{{\left({{2}\mathit{\pi}}\right)}^{2}}}{\bf k}_{\bot}^{2}\frac{{e}^{{i{\bf k}}_{\bot}\cdot{\bf x}_{\bot}}}{{k}_{\bot}}} \nonumber\\
{\times\int{\frac{{dk}_{z}}{\left({{2}\mathit{\pi}}\right)}}\left[{\frac{1}{\left({{k}_{z}^{2}{-}{\mathit{\alpha}}^{2}}\right)}{-}\frac{1}{\left({{k}_{z}^{2}{-}{\mathit{\beta}}^{2}}\right)}}\right]{e}^{{ik}_{z}z}}  \nonumber\\
+ {\frac{q}{{2}\mathit{\mu}{v}}\int{\frac{{d{\bf k}}_{\bot}}{{\left({{2}\mathit{\pi}}\right)}^{2}}}\frac{{e}^{{i{\bf k}}_{\bot}\cdot{\bf x}_{\bot}}}{{k}_{\bot}}} \nonumber\\
{\times\int{\frac{{dk}_{z}}{\left({{2}\mathit{\pi}}\right)}}\left[{\frac{{k}_{z}^{2}}{\left({{k}_{z}^{2}{-}{\mathit{\alpha}}^{2}}\right)}{-}\frac{{k}_{z}^{2}}{\left({{k}_{z}^{2}{-}{\mathit{\beta}}^{2}}\right)}}\right]{e}^{{ik}_{z}z}}, \label{qED75-d}
\end{eqnarray}
where ${\mathit{\alpha}}^{2}{=}{-}{\bf k}_{\bot}^{2}{+}\mathit{\mu}{vk}_{\bot}$ and $
{\mathit{\beta}}^{2}{=}{-}{\bf k}_{\bot}^{2}{-}\mathit{\mu}{vk}_{\bot}$.
It should be further noted that
\begin{eqnarray}
\int{\frac{{dk}_{z}}{\left({{2}\mathit{\pi}}\right)}}\left[{\frac{1}{\left({{k}_{z}^{2}{-}{\mathit{\alpha}}^{2}}\right)}{-}\frac{1}{\left({{k}_{z}^{2}{-}{\mathit{\beta}}^{2}}\right)}}\right]{e}^{{ik}_{z}z} \nonumber\\
= \frac{1}{2}\left\{{\frac{{e}^{{-}\sqrt{{k}_{\bot}^{2}{-}\mathit{\mu}{vk}_{\bot}}{z}}}{\sqrt{{k}_{\bot}^{2}{-}\mathit{\mu}{vk}_{\bot}}}{-}\frac{{e}^{{-}\sqrt{{k}_{\bot}^{2}{+}\mathit{\mu}{vk}_{\bot}}{z}}}{\sqrt{{k}_{\bot}^{2}{+}\mathit{\mu}{vk}_{\bot}}}}\right\},   \label{qED75-e}
\end{eqnarray}
and
\begin{eqnarray}
\int{\frac{{dk}_{z}}{\left({{2}\mathit{\pi}}\right)}}\left[{\frac{{k}_{z}^{2}}{\left({{k}_{z}^{2}{-}{\mathit{\alpha}}^{2}}\right)}{-}\frac{{k}_{z}^{2}}{\left({{k}_{z}^{2}{-}{\mathit{\beta}}^{2}}\right)}}\right]{e}^{{ik}_{z}z} \nonumber\\
{{=}{-}\frac{1}{2}\sqrt{{k}_{\bot}^{2}{-}\mathit{\mu}{vk}_{\bot}}{e}^{{-}\sqrt{{k}_{\bot}^{2}{-}\mathit{\mu}{vk}_{\bot}}}} \nonumber\\
{{+}\frac{1}{2}\sqrt{{k}_{\bot}^{2}{+}\mathit{\mu}{vk}_{\bot}}{e}^{{-}\sqrt{{k}_{\bot}^{2}{+}\mathit{\mu}{vk}_{\bot}}}}.  \label{qED75-f}
\end{eqnarray}

Finally, making use of the preceding results, we can write the electric field as the sum of two parts: 
\begin{equation}
{E_i} = \frac{q}{{8\pi \mu v}}{\partial _i}{\Phi ^{\left( 1 \right)}} + \frac{q}{{4\pi \mu v}}{\partial _i}{\Phi ^{\left( 2 \right)}}, \label{qED80}
\end{equation}
where
\begin{eqnarray}
{\Phi ^{\left( 1 \right)}} &=& \int_0^\infty  {d{k_ \bot }} k_ \bot ^2{J_0}\left( {{k_ \bot }|{{\bf x}_ \bot }|} \right)\frac{{{e^{ - \sqrt {k_ \bot ^2 - \mu v{k_ \bot }} z}}}}{{{{\left( {k_ \bot ^2 - \mu v{k_ \bot }} \right)}^{\frac{1}{2}}}}} \nonumber\\
 &-& \int_0^\infty  {d{k_ \bot }} k_ \bot ^2{J_0}\left( {{k_ \bot }|{{\bf x}_ \bot }|} \right)\frac{{{e^{ - \sqrt {k_ \bot ^2 + \mu v{k_ \bot }} z}}}}{{{{\left( {k_ \bot ^2 + \mu v{k_ \bot }} \right)}^{\frac{1}{2}}}}}, \nonumber\\  \label{qED85}
\end{eqnarray}
and
\begin{eqnarray}
{\Phi ^{\left( 2 \right)}} &=& \frac{1}{2}\int_0^\infty  {d{k_ \bot }} {J_0}\left( {{k_ \bot }|{{\bf x}_ \bot }|} \right)\frac{{{e^{ - \sqrt {k_ \bot ^2 + \mu v{k_ \bot }} z}}}}{{{{\left( {k_ \bot ^2 + \mu v{k_ \bot }} \right)}^{ - \frac{1}{2}}}}} \nonumber\\
&-& \frac{1}{2}\int_0^\infty  {d{k_ \bot }} {J_0}\left( {{k_ \bot }|{{\bf x}_ \bot }|} \right)\frac{{{e^{ - \sqrt {k_ \bot ^2 - \mu v{k_ \bot }} z}}}}{{{{\left( {k_ \bot ^2 - \mu v{k_ \bot }} \right)}^{ - \frac{1}{2}}}}}. \nonumber\\ \label{qED90}
\end{eqnarray}
Here ${J_0}\left( {{k_ \bot }|{{\bf x}_ \bot }|} \right)$ is a Bessel function of the first kind, where ${k_ \bot }$ and ${{\bf x}_ \bot }$  denote the momentum component and coordinate component perpendicular to $ {\bf v}$. To get the above expressions we used $
{J}_{0}\left({x}\right){=}\frac{1}{{2}\mathit{\pi}}\mathop{\int}\nolimits_{0}\nolimits^{{2}\mathit{\pi}}{{e}^{{ix}\cos\mathit{\theta}}}{d}\mathit{\theta}$.

We now are in a position to calculate the energy interaction between static point-like sources, by using the gauge-invariant but path-dependent variables formalism. This is accomplished by making use of \cite{Pato1}
\begin{equation}
V \equiv q\left( {{{\cal A}_0}\left( {\bf 0} \right) - {{\cal A}_0}\left( {\bf y} \right)} \right), \label{qED95}
\end{equation}
where the physical scalar potential is given by
\begin{equation}
{{\cal A}_0}\left( {\bf x} \right) = \int_0^1 {d\lambda } {x^i}{E_i}\left( {\lambda {\bf x}} \right), \label{qED100}
\end{equation}
with $i=1,2,3$. We also recall that (\ref{qED100}) follows from the vector gauge-invariant field expression \cite{Pato1}
\begin{equation}
{{\cal A}_\mu }\left( x \right) \equiv {A_\mu }\left( x \right) + {\partial _\mu }\left( { - \int_\xi ^x {d{z^\mu }{A_\mu }\left( z \right)} } \right), \label{qED105}
\end{equation}
where the line integral is along a space-like path from $\xi$ to $x$, on a fixed time slice. Interestingly, these variables (\ref{qED105}) commute with the sole first class constraint (Gauss's law), showing that these fields are physical variables.

Now making use of equations (\ref{qED80}) and (\ref{qED100}), we readily find that 
\begin{equation}
 {{\cal A}_0}\left( {\bf x} \right) = \frac{q}{{8\pi \mu v}}{\Phi ^{\left( 1 \right)}}\left( {\bf x} \right) + \frac{q}{{4\pi \mu v}}{\Phi ^{\left( 2 \right)}}\left( {\bf x} \right), \label{qED110}
\end{equation}
after subtracting the self-energy terms.

From equation (\ref{qED95}), the corresponding static potential for two opposite charges located at ${\bf 0}$ and ${\bf y}$ it should be calculated.

One may gain further insight into the overall structure of the interaction energy by examining equations (\ref{qED85}) and (\ref{qED90}) in some limit. With this in mind, we shall introduce a cutoff $\Lambda$ in equations (\ref{qED85}) and (\ref{qED90}).
It may be noted here that in according to the current estimates in the literature on the
parameters associated to the $v$ Lorentz-symmetry violating operators \cite{Russell}, the product of the parameters  
$\mu$ and $v$ must be upper-bounded as follows: $\mu ,v < {10^{ - 42}}$GeV. Since we are considering our cut-off, $\Lambda$, to be very small, this means that we are confined to the regime of low-frequency electromagnetic waves 
(radio waves, for instance). Therefore, we may safely take the modulus of the  wave vector (in natural units) in the range $ k \sim {10^{ - 22}} GeV - {10^{ - 15}} GeV$. Then, $\Lambda$ must be of this order too.  As a consequence, with the values estimated 
above for $\mu$, $v$, $k$  and  $\Lambda$, we can ensure that  $\mu v{k_ \bot } \ll k_ \bot ^2,{\Lambda ^2}$.

In such a case, we can rewrite expression (\ref{qED75-b}) in the form 
\begin{equation}
\Phi  = \frac{q}{{8\pi \mu v}}\mathop {\lim }\limits_{\Lambda  \to 0} {\tilde \Phi ^{\left( 1 \right)}} + \frac{q}{{4\pi \mu v}}\mathop {\lim }\limits_{\Lambda  \to 0} {\tilde \Phi ^{\left( 2 \right)}}, \label{qED115}
\end{equation}
where
\begin{eqnarray}
{{\tilde \Phi }^{\left( 1 \right)}} &=& \int_0^\infty  {d{k_ \bot }} k_ \bot ^2{J_0}\left( {{k_ \bot }|{{\bf x}_ \bot }|} \right)\frac{{{e^{ - \sqrt {k_ \bot ^2 + {\Lambda ^2} - \mu v{k_ \bot }} z}}}}{{\sqrt {k_ \bot ^2 + {\Lambda ^2} - \mu v{k_ \bot }} }} \nonumber\\
 &-& \int_0^\infty  {d{k_ \bot }} k_ \bot ^2{J_0}\left( {{k_ \bot }|{{\bf x}_ \bot }|} \right)\frac{{{e^{ - \sqrt {k_ \bot ^2 + {\Lambda ^2} + \mu v{k_ \bot }} z}}}}{{\sqrt {k_ \bot ^2 + {\Lambda ^2} + \mu v{k_ \bot }} }}, \nonumber\\ 
\label{qED115-b}
\end{eqnarray}
and
\begin{eqnarray}
{{\tilde \Phi }^{\left( 2 \right)}} &=&  - \frac{1}{2}\int_0^\infty  {d{k_ \bot }} {J_0}\left( {{k_ \bot }|{{\bf x}_ \bot }|} \right)\frac{{{e^{ - \sqrt {k_ \bot ^2 + {\Lambda ^2} - \mu v{k_ \bot }} z}}}}{{{{\left( {k_ \bot ^2 + {\Lambda ^2} - \mu v{k_ \bot }} \right)}^{ - \frac{1}{2}}}}} \nonumber\\
 &+& \frac{1}{2}\int_0^\infty  {d{k_ \bot }} {J_0}\left( {{k_ \bot }|{{\bf x}_ \bot }|} \right)\frac{{{e^{ - \sqrt {k_ \bot ^2 + {\Lambda ^2} + \mu v{k_ \bot }} z}}}}{{{{\left( {k_ \bot ^2 + {\Lambda ^2} + \mu v{k_ \bot }} \right)}^{ - \frac{1}{2}}}}}. \nonumber\\
\label{qED115-c}
\end{eqnarray}

We shall now examine the $\mu v  \ll \Lambda$ case. As a consequence of this the $\Phi$ function reads
\begin{eqnarray}
\Phi  &=& \frac{{q}}{{4\pi }}\int_0^\infty  {d{k_ \bot }} {J_0}\left( {{k_ \bot }|{{\bf x}_ \bot }|} \right){e^{ - {k_ \bot }z}} \nonumber\\
 &+&\frac{1}{64}{q}{\mathit{\mu}}^{2}{v}^{2}{z}\mathop{\int}\nolimits_{0}\nolimits^{\infty}{{dk}_{\bot}}\frac{1}{{k}_{\bot}}{J}_{0}\left({{k}_{\bot}\left|{{\bf x}_{\bot}}\right|}\right){e}^{{-}{k}_{\bot}{z}}. \nonumber\\
\label{qED120}
\end{eqnarray}

Now, making use of equations (\ref{qED95}), (\ref{qED100}) and (\ref{qED120}), we find that the potential for two opposite charges located at $\bf 0$ and $\bf r$ takes the form
\begin{equation}
V =  - \frac{{{q^2}}}{{4\pi }}\frac{1}{r} + \frac{{{q^2}}}{{4\pi }}\frac{{{\mu ^2}{v^2}}}{{16}}z\ln \left( {\frac{{z + r}}{{2z}}} \right), \label{qED125}
\end{equation}
after subtracting divergent terms, and $r = |{\bf r}|$. Evidently, by considering the limit $\mu v \to 0$, we obtain a Coulombic potential.

\subsection{The Abelian Higgs model with a Lorentz-breaking term}

We now extend what we have done to a Lorentz-violating Higgs model. However, before
going to the derivation of the interaction potential, we shall summarize very
quickly the principal features of this model. For this purpose, we start from the
four-dimensional space-time Lagrangian density \cite{Baeta}:
\begin{eqnarray}
{\cal L} &=&  - \frac{1}{4}F_{\mu \nu }^2 + |{D_\mu }\varphi {|^2} - {m^2}|\varphi {|^2} + \lambda |\varphi {|^4} \nonumber\\
&-& \frac{\mu }{4}{\varepsilon ^{\mu \nu \kappa \lambda }}{v_\mu }{A_\nu }{F_{\kappa \lambda }}, \label{ahM01}
\end{eqnarray}
where ${D_\mu } \equiv {\partial _\mu } + ieQ{A_\mu }$. As before, $v_{\mu}$, is an arbitrary four vector
which selects a preferred direction in the space-time.
Now we recall that when the gauge symmetry is spontaneously broken by means the new vacuum $\left\langle 0 \right|\varphi \left| 0 \right\rangle  = a$, where $a = {\left( { - \frac{{{m^2}}}{{2\lambda }}} \right)^{{\raise0.5ex\hbox{$\scriptstyle 1$}\kern-0.1em/\kern-0.15em\lower0.25ex\hbox{$\scriptstyle 2$}}}}$ and ${m^2} < 0$, the corresponding effective Lagrangian density reads 
\begin{equation}
{\cal L} = - \frac{1}{4}{F_{\mu \nu }}{F^{\mu \nu }} - \frac{\mu }{4}{v_\mu }{A_\nu }{F_{\alpha \beta }}{\varepsilon ^{\mu \nu \alpha \beta }} + \frac{{{M^2}}}{2}{A_\mu }{A^\mu } - A_{0}J^{0}, \label{ahM05}
\end{equation}
where ${M^2} \equiv  2{e^2}{Q^2}{a^2}$ and $J^0$ is an external source. To get the last expression we adopted a polar parametrization and used the unitary gauge.

This new effective theory provide us with a suitable starting point to study the interaction energy. Nevertheless, 
to carry out such study one would need to restore the gauge invariance in equation (\ref{ahM05}). It is with this goal that, by making use of standard techniques for constrained systems, we find that equation (\ref{ahM05}) reduces to
\begin{equation}
{\cal L} =  - \frac{1}{4}{F_{\mu \nu }}\left( {1 + \frac{{{M^2}}}{\Delta }} \right){F^{\mu \nu }} + \frac{\mu }{2}{\varepsilon ^{i\nu \alpha \beta }}{v_i}{A_\nu }\left( {{\partial _\alpha }{A_\beta }} \right) - {A_0}{J^0}, \label{ahM10}
\end{equation}
where $\Delta  \equiv {\partial _\mu }{\partial ^\mu }$.
Notice that, for notational convenience, we have maintained $\Delta$ in equation (\ref{ahM10}), but it should be borne in mind that we are considering the static case.

With the foregoing information, we proceed to obtain the Hamiltonian. The canonical momenta are ${\Pi ^\mu } =  - \left( {1 + \frac{{{M^2}}}{\Delta }} \right){F^{0\mu }} + \frac{\mu }{2}{\varepsilon ^{i\nu 0\mu }}{v_i}{A_\nu }$, and one immediately identifies the primary constraint ${\Pi ^0} = 0$, while the remaining non-zero momenta are ${\Pi ^i} =  - \left( {1 + \frac{{{M^2}}}{\Delta }} \right){F^{0i}} + \frac{\mu }{2}{\varepsilon ^{jk0i}}{v_j}{A_k}$. The canonical Hamiltonian is now obtained in the usual way and is given by
\begin{eqnarray}
{H_C} &=& \int {{d^3}x} \left\{ { - {A_0}\left( {{\partial _i}{\Pi ^i} - \frac{\mu }{2}{\varepsilon ^{0ijk}}{v_i}{\partial _j}{A_k} 
+ {J^0}} \right)} \right\} \nonumber\\
&+&\int {{d^3}x} \left\{ { - \frac{1}{2}{F_{i0}}\left( {1 + \frac{{{M^2}}}{\Delta }} \right){F^{i0}}} \right\} \nonumber\\
&+& \int {{d^3}x} \left\{ {\frac{1}{4}{F_{ij}}\left( {1 + \frac{{{M^2}}}{\Delta }} \right){F^{ij}}} \right\}.  \label{ahM15}
\end{eqnarray}
As before, requiring the primary constraint ${\Pi ^0}$ to be preserved in time yields the secondary constraint (Gauss's law) ${\Gamma _1} \equiv {\partial _i}{\Pi ^i} - \frac{\mu }{2}{\varepsilon ^{0ijk}}{v_i}{\partial _j}{A_k} + {J^0} = 0$. Thus, the Hamiltonian is now given as 
\begin{eqnarray}
{H} &=& \int {{d^3}x} \left\{ {w\left( x \right)\left( {{\partial _i}{\Pi ^i} - \frac{\mu }{2}{\varepsilon ^{0ijk}}{v_i}{\partial _j}{A_k} + {J^0}} \right)} \right\} \nonumber\\
&+&\int {{d^3}x} \left\{ { - \frac{1}{2}{F_{i0}}\left( {1 + \frac{{{M^2}}}{\Delta }} \right){F^{i0}}} \right\} \nonumber\\
&+& \int {{d^3}x} \left\{ {\frac{1}{4}{F_{ij}}\left( {1 + \frac{{{M^2}}}{\Delta }} \right){F^{ij}}} \right\},  
\label{ahM20}
\end{eqnarray}
where, as before, $w\left( x \right) = {u_1}\left( x \right) - {A_0}\left( x \right)$.

Since our goal is to compute the static potential for the theory under consideration, we shall use the same gauge-fixing condition that was used in our preceding calculation. In view of this situation, we now write the Dirac brackets in terms of the magnetic and electric fields as
\begin{equation}
{\left\{ {{E_i}\left( x \right),{B_j}\left( y \right)} \right\}^ * } =  - {\left( {1 + \frac{{{M^2}}}{\Delta }} \right)^{ - 1}}\varepsilon _j^{ki}{\partial _k}{\delta ^{\left( 3 \right)}}\left( {{\bf x} - {\bf y}} \right), \label{ahM25-a}
\end{equation}

\begin{equation}
{\left\{ {B\left( x \right),B\left( y \right)} \right\}^ * } = 0, \label{ahM25-b}
\end{equation}

\begin{equation}
{\left\{ {{E_i}\left( x \right),{E_j}\left( y \right)} \right\}^ * } =  - {\left( {1 + \frac{{{M^2}}}{\Delta }} \right)^{ - 2}}\mu {\varepsilon _{ijk}}{v^k}{\delta ^{\left( 3 \right)}}\left( {{\bf x} - {\bf y}} \right). \label{ahM25-c}
\end{equation}

It gives rise to the following equations of motion for the magnetic and electric fields:
\begin{equation}
{\dot E_i}\left( x \right) =  - \mu {\left( {1 + \frac{{{M^2}}}{\Delta }} \right)^{ - 1}}{\varepsilon _{ijk}}{v_k}{E_j}\left( x \right) + {\varepsilon _{ijk}}{\partial _k}{B_j}\left( x \right), \label{ahM30}
\end{equation}
and
\begin{equation}
{\dot B_i}\left( x \right) =  + {\varepsilon _{jik}}{\partial _k}{E_j}\left( x \right). \label{ahM35}
\end{equation}
It follows from the above discussion that Gauss's law for the present theory reads
\begin{equation}
\left( {1 + \frac{{{M^2}}}{\Delta }} \right){\partial _i}{\Pi ^i} - \mu {v_i}{B^i} + {J^0} = 0. \label{ahM40}
\end{equation}

Again, as in the previous subsection, we shall consider static fields. Thus, we obtain
\begin{equation}
 {E_i} = {\partial _i}\Phi,  \label{ahM45}
\end{equation}
where
\begin{equation}
\Phi  = \frac{{\left( {{\nabla ^2} - {M^2}} \right)}}{{\left[ {{{\left( {{\nabla ^2} - {M^2}} \right)}^2} + {\mu ^2}{v^2}{\nabla ^2} - {\mu ^2}{{\left( {{\bf v} \cdot \nabla } \right)}^2}} \right]}}\left( { - {J^0}} \right). \label{ahM50}
\end{equation}

For ${J^0}\left( {\bf x} \right) = q{\delta ^{\left( 3 \right)}}\left( {\bf x} \right)$, the foregoing expression becomes
\begin{equation}
\Phi  = \frac{q}{{8\pi \mu v}}{\Phi ^{\left( 1 \right)}} + \frac{q}{{16\pi \mu v}}{\Phi ^{\left( 2 \right)}} + \frac{{q{M^2}}}{{8\pi \mu v}}{\Phi ^{\left( 3 \right)}},  \label{ahM55}
\end{equation}
where
\begin{eqnarray}
{\Phi ^{\left( 1 \right)}} &=& \int_0^\infty  {d{k_ \bot }} k_ \bot ^2{J_0}\left( {{k_ \bot }|{{\bf x}_ \bot }|} \right)\frac{{{e^{ - \sqrt {k_ \bot ^2 + {M^2} - \mu v{k_ \bot }} z}}}}{{{{\left( {k_ \bot ^2 + {M^2} - \mu v{k_ \bot }} \right)}^{\frac{1}{2}}}}} \nonumber\\
&-& \int_0^\infty  {d{k_ \bot }} k_ \bot ^2{J_0}\left( {{k_ \bot }|{{\bf x}_ \bot }|} \right)\frac{{{e^{ - \sqrt {k_ \bot ^2 + {M^2} + \mu v{k_ \bot }} z}}}}{{{{\left( {k_ \bot ^2 + {M^2} + \mu v{k_ \bot }} \right)}^{\frac{1}{2}}}}}, \nonumber\\
\label{ahM60}
\end{eqnarray}
\begin{eqnarray}
{\Phi ^{\left( 2 \right)}} &=& \int_0^\infty  {d{k_ \bot }} {J_0}\left( {{k_ \bot }|{{\bf x}_ \bot }|} \right)\frac{{{e^{ - \sqrt {k_ \bot ^2 + {M^2} + \mu v{k_ \bot }} z}}}}{{{{\left( {k_ \bot ^2 + {M^2} + \mu v{k_ \bot }} \right)}^{ - \frac{1}{2}}}}} \nonumber\\
 &-& \int_0^\infty  {d{k_ \bot }} {J_0}\left( {{k_ \bot }|{{\bf x}_ \bot }|} \right)\frac{{{e^{ - \sqrt {k_ \bot ^2 + {M^2} - \mu v{k_ \bot }} z}}}}{{{{\left( {k_ \bot ^2 + {M^2} - \mu v{k_ \bot }} \right)}^{ - \frac{1}{2}}}}}, \nonumber\\
\label{ahM65}
\end{eqnarray}
and
\begin{eqnarray}
{\Phi ^{\left( 3 \right)}} &=& \int_0^\infty  {d{k_ \bot }} {J_0}\left( {{k_ \bot }|{{\bf x}_ \bot }|} \right)\frac{{{e^{ - \sqrt {k_ \bot ^2 + {M^2} - \mu v{k_ \bot }} z}}}}{{{{\left( {k_ \bot ^2 + {M^2} - \mu v{k_ \bot }} \right)}^{\frac{1}{2}}}}} \nonumber\\
&-& \int_0^\infty  {d{k_ \bot }} {J_0}\left( {{k_ \bot }|{{\bf x}_ \bot }|} \right)\frac{{{e^{ - \sqrt {k_ \bot ^2 + {M^2} + \mu v{k_ \bot }} z}}}}{{{{\left( {k_ \bot ^2 + {M^2} + \mu v{k_ \bot }} \right)}^{\frac{1}{2}}}}}. \nonumber\\
\label{ahM70}
\end{eqnarray}

We now have all the information required to compute the potential energy for static charges in this theory. Thus, by employing equation (\ref{qED100}), the gauge-invariant scalar potential may be rewritten as
\begin{equation}
{{\cal A}_0}\left( {\bf x} \right) = \frac{q}{{8\pi \mu v}}{\Phi ^{\left( 1 \right)}} + \frac{q}{{16\pi \mu v}}{\Phi ^{\left( 2 \right)}} + \frac{{q{M^2}}}{{8\pi \mu v}}{\Phi ^{\left( 3 \right)}}, \label{ahM75}
\end{equation}
after subtracting the self-energy terms.

As was explained before, from equation (\ref{qED95}), the corresponding static potential for two opposite charges located at ${\bf 0}$ and ${\bf r}$ it should be calculated.

However, following our earlier line of argument, we shall now consider the background small compared with the mass term $({\mu ^2}{{\bf v}^2} \ll {M^2})$. Accordingly, expression (\ref{ahM55}) can be simplified
\begin{equation}
\Phi  = \frac{q}{{8\pi \mu v}}{\Phi ^{\left( 1 \right)}} + \frac{q}{{16\pi \mu v}}{\Phi ^{\left( 2 \right)}} + \frac{{q{M^2}}}{{8\pi \mu v}}{\Phi ^{\left( 3 \right)}}, \label{ahM80}
\end{equation}
where
\begin{eqnarray}
{\Phi ^{\left( 1 \right)}} &=& \mu v\int_0^\infty  {d{k_ \bot }} k_ \bot ^3{J_0}\left( {{k_ \bot }|{{\bf x}_ \bot }|} \right)\frac{{{e^{ - \sqrt {k_ \bot ^2 + {M^2}} z}}}}{{{{\left( {k_ \bot ^2 + {M^2}} \right)}^{\frac{3}{2}}}}} \nonumber\\
 &+& \frac{{30{\mu ^3}{v^3}}}{{48}}\int_0^\infty  {d{k_ \bot }} k_ \bot ^5{J_0}\left( {{k_ \bot }|{{\bf x}_ \bot }|} \right)\frac{{{e^{ - \sqrt {k_ \bot ^2 + {M^2}} z}}}}{{{{\left( {k_ \bot ^2 + {M^2}} \right)}^{\frac{7}{2}}}}}, \nonumber\\\label{ahM80-a}
\end{eqnarray}
\begin{eqnarray}
{\Phi ^{\left( 2 \right)}} &=& \mu v\int_0^\infty  {d{k_ \bot }} {k_ \bot }{J_0}\left( {{k_ \bot }|{{\bf x}_ \bot }|} \right)\frac{{{e^{ - \sqrt {k_ \bot ^2 + {M^2}} z}}}}{{{{\left( {k_ \bot ^2 + {M^2}} \right)}^{\frac{1}{2}}}}}     \nonumber\\
&+& \frac{{{\mu ^3}{v^3}}}{8}\int_0^\infty  {d{k_ \bot }} k_ \bot ^3{J_0}\left( {{k_ \bot }|{{\bf x}_ \bot }|} \right)\frac{{{e^{ - \sqrt {k_ \bot ^2 + {M^2}} z}}}}{{{{\left( {k_ \bot ^2 + {M^2}} \right)}^{\frac{5}{2}}}}}, \nonumber\\
 \label{ahM80-b}
\end{eqnarray}
and
\begin{eqnarray}
{\Phi ^{\left( 3 \right)}} &=& \mu v\int_0^\infty  {d{k_ \bot }} {k_ \bot }{J_0}\left( {{k_ \bot }|{{\bf x}_ \bot }|} \right)\frac{{{e^{ - \sqrt {k_ \bot ^2 + {M^2}} z}}}}{{{{\left( {k_ \bot ^2 + {M^2}} \right)}^{\frac{3}{2}}}}}      \nonumber\\
&+& \frac{{5{\mu ^3}{v^3}}}{8}\int_0^\infty  {d{k_ \bot }} k_ \bot ^3{J_0}\left( {{k_ \bot }|{{\bf x}_ \bot }|} \right)\frac{{{e^{ - \sqrt {k_ \bot ^2 + {M^2}} z}}}}{{{{\left( {k_ \bot ^2 + {M^2}} \right)}^{\frac{7}{2}}}}}.  \nonumber\\
\label{ahM80-c}
\end{eqnarray}

After some further manipulations, equation (\ref{ahM80}) reduces to
\begin{eqnarray}
\Phi  &=& \frac{{3q}}{{16\pi }}\frac{{{e^{ - Mr}}}}{r} - \frac{q}{{8\pi }}\frac{{11}}{{16}}{\mu ^2}{v^2}\left( {r + \frac{1}{M} + \frac{{M{z^2}}}{2}} \right)\frac{{{e^{ - Mr}}}}{3} \nonumber\\
&-& \frac{q}{{8\pi }}\frac{{11}}{{16}}{\mu ^2}{v^2}\left\{ {z\left( {1 + \frac{{{M^2}{z^2}}}{2}} \right){K_0}\left( {M|{{\bf x}_ \bot }|} \right)} \right\} \nonumber\\
&+& \frac{q}{{8\pi }}\frac{{11}}{{16}}{\mu ^2}{v^2}\left\{ {z\left( {1 - \frac{{{M^2}{z^2}}}{6}} \right)\int_1^{\frac{r}{{|{{\bf x}_ \bot }|}}} {du\frac{{{e^{ - M|{{\bf x}_ \bot }|u}}}}{{\sqrt {{u^2} - 1} }}} } \right\} \nonumber\\
&+& \frac{q}{{8\pi }}\frac{{11}}{{16}}{\mu ^2}{v^2}\left\{ {\frac{{Mz|{{\bf x}_ \bot }|}}{2}{K_1}\left( {M|{{\bf x}_ \bot }|} \right)} \right\} \nonumber\\
&-&\frac{q}{{8\pi }}\frac{{11}}{{16}}{\mu ^2}{v^2}{M^2}\frac{{z|{{\bf x}_ \bot }{|^2}}}{2}\int_1^{\frac{r}{{|{{\bf x}_ \bot }|}}} {du{e^{ - M|{{\bf x}_ \bot }|u}}\sqrt {{u^2} - 1} }. \nonumber\\
\label{ahM85}
\end{eqnarray}
Here $K_{0}$ and $K_{1}$ are modified Bessel functions.

Once again, following our earlier procedure, we get the interaction energy as
\begin{eqnarray}
V &=&  - \frac{{3{q^2}}}{{16\pi }}\frac{{{e^{ - Mr}}}}{r} + \frac{{{q^2}}}{{8\pi }}\frac{{11}}{{16}}{\mu ^2}{v^2}\left( {r + \frac{1}{M} + \frac{{M{z^2}}}{2}} \right)\frac{{{e^{ - Mr}}}}{3} \nonumber\\
&+& \frac{{{q^2}}}{{8\pi }}\frac{{11}}{{16}}{\mu ^2}{v^2}\left\{ {z\left( {1 + \frac{{{M^2}{z^2}}}{2}} \right){K_0}\left( {M|{{\bf x}_ \bot }|} \right)} \right\} \nonumber\\
&-& \frac{{{q^2}}}{{8\pi }}\frac{{11}}{{16}}{\mu ^2}{v^2}\left\{ {z\left( {1 - \frac{{{M^2}{z^2}}}{6}} \right)\int_1^{\frac{r}{{|{{\bf x}_ \bot }|}}} {du\frac{{{e^{ - M|{{\bf x}_ \bot }|u}}}}{{\sqrt {{u^2} - 1} }}} } \right\} \nonumber\\
&-& \frac{{{q^2}}}{{8\pi }}\frac{{11}}{{16}}{\mu ^2}{v^2}\left\{ {\frac{{Mz|{{\bf x}_ \bot }|}}{2}{K_1}\left( {M|{{\bf x}_ \bot }|} \right)} \right\} \nonumber\\
&+&\frac{{{q^2}}}{{8\pi }}\frac{{11}}{{16}}{\mu ^2}{v^2}{M^2}\frac{{z|{{\bf x}_ \bot }{|^2}}}{2}\int_1^{\frac{r}{{|{{\bf x}_ \bot }|}}} {du{e^{ - M|{{\bf x}_ \bot }|u}}\sqrt {{u^2} - 1} }. \nonumber\\
\label{ahM90}
\end{eqnarray}
In this way our calculation shows new corrections to the Yukawa potential. Evidently, by considering the limit $\mu v \to 0$, we obtain a Proca-like theory. Finally, it is worth mentioning that the previous calculation generalizes the results presented in Ref. \cite{Casana3}, where electric and magnetic fields were calculated for the special case where the background $v$ and $\bf x$ are parallel.

\section{Final Remarks}

Finally, by exploiting the gauge-invariant but path-dependent variables formalism, we have addressed the confinement versus screening issue for both  electrodynamics of chiral matter and an Abelian Higgs-like model in $3+1$ dimensions. An important feature of this framework is a correct identification of physical degrees of freedom for understanding the physics hidden in gauge theories. As a consequence, in the case of electrodynamics of chiral matter and a purely space-like vector, $v^{\mu}$, we have obtained a logarithmic correction to the usual static Coulomb potential. On the other hand, in the case of the Abelian Higgs model with a Lorentz-breaking term and a purely space-like vector, $v^{\mu}$, the static potential displays new corrections to the Yukawa potential.\\

\section{ACKNOWLEDGMENTS}
P. G. was partially supported by Fondecyt (Chile) grant 1180178 and by Proyecto Basal FB0821.


\begin{thebibliography}{}

\bibitem{Peccei-Quinn} R. D. Peccei and H. R. Quinn, Phys. Rev. Lett. {\bf 38}, 1440 (1977).

\bibitem{Weinberg} S. Weinberg, Phys. Rev. Lett. {\bf 40}, 223 (1978).

\bibitem{Wilczek1} F. Wilczek, Phys. Rev. Lett. {\bf 40}, 279 (1978).

\bibitem{Witten} E. Witten, Phys. Lett. B {\bf 86}, 283 (1979). 

\bibitem{Rosenberg} M. Franz and G. Rosenberg, Phys. Rev. B {\bf 82}, 035105 (2010).

\bibitem{Franz} M. M. Vazifeh and M. Franz, Rhys. Rev. D {\bf 82}, 233103 (2010).

\bibitem{Qi} X. L. Qi, T. L. Hughes and S. -C. Zhang, Science {\bf 323}, 1184 (2009).

\bibitem{Essin} A. M. Essin, J. E. Moore and D. Vanderbilt, Phys. Rev. Lett. {\bf 102}, 146805 (2009.)

\bibitem{Kane} M. Hasan and C. Kane, Rev. Mod. Phys. {\bf 82}, 3045 (2010).

\bibitem{Cao} Z. Qiu, G. Cao and Xu-Guang Huang, Phys. Rev. D {\bf 95}, 036002 (2017).

\bibitem{Kharzeev1} D. E. Kharzeev, L. D. McLerran and H. J. Warringa, Nucl. Phys. A {\bf 803}, 227 (2008).

\bibitem{Kharzeev2} D. E. Kharzeev, Prog. Part. Nucl. Phys. {\bf 75}, 133 (2014).

\bibitem{Tuchin} K. Tuchin, Phys. Rev. C {\bf 91}, 064902 (2015).

\bibitem{Hayata} T. Hayata, Phys. Rev. B {\bf 97}, 205102 (2018). 

\bibitem{Haldane} F. D. M. Haldane, Phys. Rev. Lett. {\bf 93}, 206602 (2004).

\bibitem{Huang} Xu-Guang Huang, Rep. Prog. Phys. {\bf 79}, 076302 (2016).

\bibitem{Jackiw} S. M. Carroll, G. B. Field and R. Jackiw, Phys. Rev. D {\bf 41}, 1231 (1990).

\bibitem{Kostelecky} D. Colladay and V. A. Kostelecky, Phys. Rev. D {\bf 58}, 116002 (1998).

\bibitem{Liberati} S. Liberati, Class. Quantum Grav. {\bf 30}, 133001 (2013).

\bibitem{VAKostelecky} V.~A.~Kostelecky,
``Proceedings, 7th Meeting on CPT and Lorentz Symmetry (CPT 16) : Bloomington, Indiana, USA, June 20-24, (2016).

\bibitem{Deser} S. Deser, R. Jackiw and S. Templeton, Ann. Phys. (N.Y.) {\bf 140}, 372 (1982).

\bibitem{Dunne}  G. Dunne, hep-th/9902115.

\bibitem{Khare}  A. Khare, {\it Fractional Statistics and Quantum Theory} (World Scientific, 1998).

\bibitem{Banerjee} R. Banerjee, A. Chatterjee and V. V. Sreedhar, Ann. Phys. (N.Y.) {\bf 222}, 254 (1993).

\bibitem{Wilczek2} F. Wilczek, Phys. Rev. Lett. {\bf 58}, 1799 (1987).

\bibitem{Baeta} A. P. Baeta Scarpelli, H. Belich, J. L. Boldo and J. Helay\"el-Neto, Phys. Rev. D {\bf 67}, 085021 (2003).

\bibitem{Casana1} R. Casana, M. M. Ferreira Jr. and A. L. Mota, Annals of Physics, {\bf 375}, 179 (2016).

\bibitem{Casana2} R. Casana, M. M. Ferreira Jr., E. da Hora and A. B. F. Neves, Eur. Phys. J. C {\bf 74}, 3064 (2014).

\bibitem{Pato1}  P. Gaete, Z. Phys. C{\bf 76}, 355 (1997).

\bibitem{Pato2} P.~Gaete and I.~Schmidt, Phys.\ Rev.\ D {\bf 64}, 027702 (2001).

\bibitem{Russell} V. A. Kostelecky and N. Russell, Rev. Mod. Phys. {\bf 83}, 11 (2011). 

\bibitem{Casana3} R. Casana, M. M. Ferreira Jr. and C. E. H. Santos, Phys. Rev. D {\bf 78} 025030 (2008).
\end{thebibliography}
\end{document}